\documentclass[aps,prb,twocolumn,twocolumn,showpacs,amsfonts]{revtex4-1}

\usepackage{amsmath}
\usepackage{graphicx}

\newcommand{\beq}{\begin{equation}}
\newcommand{\eeq}{\end{equation}}
\newcommand{\beqarray}{\begin{eqnarray}}
\newcommand{\eeqarray}{\end{eqnarray}}

\begin{document}

\title{Doping dependence of antiferromagnetism in models of the pnictides}

\author{Jacob Schmiedt}
\email{jacob\_alexander.schmiedt@tu-dresden.de}
\affiliation{Institut f\"{u}r Theoretische Physik, Technische Universit\"{a}t
  Dresden, 01062 Dresden, Germany}

\author{P. M. R. Brydon} 
\email{brydon@theory.phy.tu-dresden.de}
\affiliation{Institut f\"{u}r Theoretische Physik, Technische Universit\"{a}t
  Dresden, 01062 Dresden, Germany}

\author{Carsten Timm}
\email{carsten.timm@tu-dresden.de}
\affiliation{Institut f\"{u}r Theoretische Physik, Technische Universit\"{a}t
  Dresden, 01062 Dresden, Germany}

\date{February 24, 2012}

\begin{abstract}
We study the doping dependence of the spin-density-wave (SDW) state in four
models of the 1111 pnictides. The random-phase approximation is used to
determine the ordering temperature and the ordering vector as
functions of doping, and to evaluate the contribution of the
various orbitals to the SDW instability. In addition to the
usual assumption of orbitally rotation-invariant interactions, we consider
the effect of reduced interactions involving the $xy$ orbital, which are
anticipated by crystal-structure considerations.
We find that changing the relative strength of the interaction in the
$xy$ orbital tunes the system between different nesting
instabilities leading to similar SDW order. We identify two
models as showing reasonable agreement with experiments, while the other two
display significant discrepancies, and discuss the underlying differences
between the models. 
\end{abstract}


\maketitle

\section{Introduction}\label{intro}

The discovery of superconductivity in  the iron pnictides has triggered
intense research activity striving to understand this diverse class of
materials. Two pnictide families have attracted particular interest due
to their high critical temperatures, namely the 1111 compounds\cite{kamihara}
{\it R}FeAsO and the 122 
compounds\cite{rotter} $A{\rm Fe_2As_2}$ ({\it R} and {\it A} are
rare-earth and 
alkaline-earth elements, respectively). Both system are antiferromagnets at
parent-compound filling, and must be chemically doped or subjected to
significant pressure to become superconducting. The close proximity of
antiferromagnetic (AFM) and superconducting states in the phase diagram
evidences an intimate connection between the magnetic and the
superconducting phases.\cite{zhao,huang}

There is considerable experimental evidence that the 1111 and the 122
families display a metallic spin-density-wave (SDW)
state.\cite{lumsden,sebastian,yi,shimojima,mcguire,liu,dong}  Furthermore,
spectroscopic measurements indicate at most intermediate correlation
strengths.\cite{drechsler,yang} This suggests that the
microscopic origin of the AFM phase can be understood within a weak-coupling
picture, which is consistent with {\it ab initio} calculations 
identifying the
nesting between the electron and hole Fermi pockets as responsible for
the SDW.\cite{zhang,singh,mazin}

The nesting picture of the magnetism implies a strong doping dependence of the
AFM phase. It is well established that the SDW critical temperature $T_{c}$ 
of the 1111 compounds decreases upon electron doping.\cite{lumsden} 
In contrast, very few experiments with hole-doped 1111 compounds have been
reported, and the magnetic behavior remains
obscure.\cite{mu,mu2,mu3,bernhard,wen} The resistivity anomaly characteristic
of the onset of SDW in the parent compounds becomes much less pronounced upon
hole doping in ${\rm La}_{1-x}{\rm Sr}_x{\rm
FeAsO}$,\cite{wen,mu3}  ${\rm Pr}_{1-x}{\rm Sr}_x{\rm FeAsO}$,\cite{mu} and
${\rm Pr}_{1-x}{\rm Ca}_x{\rm FeAsO}$,\cite{mu2} but it is unclear whether
this corresponds to a suppression of magnetic order. 
A more accurate probe of the antiferromagnetism is given by $\mu$SR
measurements, which in ${\rm Pr}_{1-x}{\rm Sr}_x{\rm FeAsO}$ suggests that
mesoscopic phase separation allows a substantial fraction of the system
to remain antiferromagnetic up to $x=0.2$, although it is not known if this is
intrinsic.\cite{bernhard} The experimental situation is much clearer in the 
122 family, e.g., \emph{electron}-doping rapidly suppresses the SDW in ${\rm
Ba(Fe}_{1-x}{\rm Co}_x)_2{\rm As}_2$, the SDW is already absent for
$x=0.05$,\cite{chu} while for the \emph{hole}-doped ${\rm Ba}_{1-x}{\rm
K}_x{\rm Fe_2As}_2$, and the SDW $T_c$ decreases more slowly and vanishes at
$0.3<x<0.4$, corresponding to $0.15$--$0.2$ holes per single-Fe unit
cell.\cite{chen,rotter2009} It is interesting to see whether this asymmetric
doping dependence of the AFM order can be replicated by theoretical models.  

The complicated structure of the Fermi surface of the pnictides, which consists 
of two electron pockets and up to three hole pockets of various orbital
character, allows for a number of
different nesting instabilities that may compete with or reinforce each
other. Previous works have identified two instabilities
that may be responsible for the characteristic $(\pi,0)$ order:
Many authors\cite{vorontsov,cvetkovic,eremin,yu,brydonEI,knolle2010,
QPinterference,brydon,daghofer,fernandes} 
identify the most important nesting as that between the
hole pockets around the $\Gamma$ point and the electron pocket at the X point
of the single-Fe Brillouin zone, which implies a dominant role of the 
Fe $3d$ $xz$ and $yz$ orbitals in the formation of the SDW state. On
the other hand, the role of nesting between the pockets at
the Y and M points has also been
emphasized,\cite{ikeda2008,arita,kuroki2,ikeda,ikeda2010}
where the key contribution comes from the Fe $3d$ $xy$
orbital. The two nesting instabilities are not mutually
exclusive, and appear at similar doping levels. However,
the number and location of the hole pockets involved in the magnetic
instability nevertheless has important consequences for the spin wave
spectrum\cite{knolle2010,knolle2011} and the realized commensurate SDW
order.\cite{brydonArXiv}  
One of the  goals of this work is therefore to examine the
relative importance of the two nesting instabilties across the phase
diagram.

The most popular starting point for the theoretical description of the
pnictides are extended Hubbard models possessing orbital degrees of
freedom. These models typically include up to five of the Fe $3d$
orbitals\cite{raghu,daghofer,yu,graser,arita,kuroki,ikeda,miyake,calderon,Luo}
or also the As $4p$ orbitals.\cite{yanagi,Schickling} Since the former
dominate the Fermi 
surface,\cite{singh,boeri,vildosola} models constructed in terms of the five
Fe $3d$ 
orbitals are expected to give a good account of the physics while keeping the
parameter space manageable. In the study of these models it is common to
assume orbitally rotation-invariant
interactions.~\cite{ikeda2008,ikeda2010,arita,daghofer,yu,kuroki,graser,ikeda,
brydon,Luo,kuroki2,kaneshita,bascones,yanagi,Schickling} 
This symmetry is however broken by the underlying crystal structure of
the FeAs lattice. Furthermore, upon integrating out the As $4p$ orbitals to
obtain an effective Fe $3d$ model, the maximally localized Wannier functions of 
the $3d$ orbitals acquire a significant As $4p$ component, and
are consequently expanded compared to the atomic
limit.\cite{vildosola,miyake} The $xy$ orbital has the
greatest overlap with the relevant $4p$ states and hence the greatest extent,
which implies a weaker
intraorbital interaction strength. Conversely, the $x^2-y^2$ and $3z^2-r^2$
orbitals have only small overlap with the $4p$ orbitals and are thus expected to
remain rather compact, with stronger interaction potentials. The degenerate
$xz$ and $yz$ orbitals lie in between these extremes. Miyake \textit{et
al.}\cite{miyake} have
confirmed the expected hierarchy of 
interaction strengths within a constrained random-phase approximation
(cRPA). The $xz$, $yz$, and $xy$ orbitals make up 
the bulk of the Fermi surface,~\cite{boeri} and so they are most important
when considering 
the origin of the SDW. The relatively weaker interaction strength on the $xy$
orbital might therefore be expected to play a significant role in selecting
the leading nesting instability. Investigating this aspect of the physics is
the second major goal of our paper.

In order to answer these questions we consider the doping dependence of the 
magnetic order in four different five-orbital models of LaFeAsO, proposed by
Kuroki \textit{et al.},\cite{kuroki} Graser \textit{et al.},\cite{graser} Ikeda
\textit{et al.},\cite{ikeda} and Calder\'{o}n \textit{et al.}\cite{calderon}
All four models are based on \textit{ab initio} band structures. Kuroki
\textit{et al.}\ and Ikeda \textit{et al.}\ obtain tight-binding models from
LDA calculations employing maximally localized Wannier functions, while Graser
\textit{et al.}\ fit a Slater-Koster tight-binding model to a GGA band
structure. Calder\'{o}n \textit{et al.}\ propose a Slater-Koster model
containing a limited number of free parameters, which were chosen to best
reproduce the LDA band structure.
We employ the random-phase approximation (RPA) to
calculate the static spin susceptibility in the paramagnetic phase, and
examine it for divergences as the temperature and doping are varied. This
allows us to determine the limits of the paramagnetic state in an unbiased way
as we are able to identify ordered states with large unit cells that would not
be accessible by the usual mean-field
approaches.\cite{brydon,yu,kaneshita,daghofer,bascones,Luo} Further insight
into the nesting mechanisms in the models is obtained by decomposing
the paramagnetic RPA spin susceptibility into its orbital components close to
the critical temperature. The comparison of the magnetic phase diagrams
of these models is the final major goal of our work. Not only does this
allow us to better understand the mechanisms for antiferromagnetism in the
pnictides, but also it helps to identify the most realistic 
models for this system.

The paper is organized as follows. In Sec.\ \ref{theory} we introduce the
model Hamiltonian and outline the calculation of the RPA spin
susceptibility. We proceed to conduct a systematic analysis of the four
different models in Sec.\ \ref{results}. This is followed in
Sec.\ \ref{discussion} by a discussion of our results and the implications for
the understanding of the pnictides. We conclude with a summary in
Sec.\ \ref{summary}.  

\section{Theory}\label{theory}

\subsection{Model}\label{model}

We start with a tight binding Hamiltonian that describes the non-interacting
five-orbital system 
\begin{align}
H_0 = \sum_{\bf k}\sum_\sigma\sum_{\nu,\mu}T_{\nu,\mu}({\bf k})d^\dagger_{{\bf
    k},\nu,\sigma}d_{{\bf k},\mu,\sigma}, 
\end{align}
where $d^\dagger_{{\bf k},\nu,\sigma}$ ($d_{{\bf k},\nu,\sigma}$) is the
creation (annihilation) operator for a spin $\sigma$ electron of momentum
${\bf k}$ in the orbital $\nu$. The values of $T_{\nu,\mu}$ for the different
models are provided in Refs.\
\onlinecite{kuroki},~\onlinecite{graser},~\onlinecite{ikeda},
and~\onlinecite{calderon}. {Keeping} only local terms in the
interaction Hamiltonian, we have
\beqarray
H_{\rm int}&=&\sum_{\bf
  i}\sum_{\nu\mu}\sum_{\sigma\sigma'}\Big[ U_{\nu\mu}d_{{\bf
    i},\nu,\sigma}^{\dagger}d_{{\bf i},\mu,\sigma'}^{\dagger}d_{{\bf
    i},\mu,\sigma'}d_{{\bf i},\nu,\sigma}  \nonumber \\ 
&& {}+ J_{\nu\mu} \Big(d_{{\bf i},\nu,\sigma}^{\dagger}d_{{\bf
    i},\mu,\sigma'}^{\dagger}d_{{\bf i},\nu,\sigma'}d_{{\bf
    i},\mu,\sigma} \nonumber \\ 
&& {}+ d_{{\bf i},\nu,\sigma}^{\dagger}d_{{\bf
    i},\nu,\sigma'}^{\dagger}d_{{\bf i},\mu,\sigma}d_{{\bf
    i},\mu,\sigma}\Big)\Big],  
\label{orb-dep}
\eeqarray
where the index ${\bf i}$ stands for the lattice site, $\nu$ and $\mu$ stand
for the orbitals, and $\sigma$ and $\sigma'$ denote the
spin. Equation~(\ref{orb-dep}) is
usually~\cite{ikeda2008,ikeda2010,arita,daghofer,yu,kuroki,graser,ikeda,brydon,
Luo,kuroki2,kaneshita,bascones,yanagi}
simplified by assuming orbital-independent interactions for which we then have
\beqarray
H_{\rm int} &=& U\sum_{\bf i}\sum_\nu n_{{\bf
i},\nu,\uparrow}n_{{\bf
    i},\nu,\downarrow}  \nonumber \\ 
&& {}+ \frac{1}{4}(2U-5J)\sum_{\bf
  i}\sum_{\nu\neq\mu}\sum_{\sigma,\sigma'}n_{{\bf
    i},\nu,\sigma}n_{{\bf i},\mu,\sigma'}\nonumber \\  
&& {}- J\sum_{\bf i}\sum_{\nu\neq\mu} {\bf S}_{{\bf i},\nu}\cdot{\bf S}_{{\bf
    i},\mu} \nonumber \\
&&+ J\sum_{\bf i}\sum_{\nu\neq\mu}d^\dagger_{{\bf 
    i},\nu\uparrow}d^\dagger_{{\bf i},\nu\downarrow}d_{{\bf
    i},\mu\downarrow}d_{{\bf i},\mu\uparrow}.
\label{orb-ind}
\eeqarray
This is obtained from Eq.\ (\ref{orb-dep}) by setting $U_{\nu\nu}=U$,
$J_{\nu\mu}=J$ and $U_{\nu\mu}=\frac{1}{4}(2U-5J)$ if $\nu\neq\mu$. The latter
choice implies invariance of the interaction Hamiltonian under rotations in
orbital space.\cite{oles,dagotto} The spin operators are expressed in terms of
the creation and annihilation operators as ${\bf
S}_{i,\nu}=\frac{1}{2}\sum_{\sigma\sigma'}d^{\dagger}_{i,\nu,\sigma}{\pmb
\sigma}_{\sigma\sigma'}d_{i,\nu\sigma'}$, where ${\pmb \sigma}$ is the
vector of Pauli matrices. The ratio between the local Coulomb interaction
and the Hund's rule coupling is set to $U/J=4$. 
For our calculations we adopt the standard assumption that doping does not
change the interaction and band structure parameters in the Hamiltonian.

Although the invariance of the interaction Hamiltonian under orbital
rotations is a common
assumption,~\cite{ikeda2008,ikeda2010,arita,daghofer,yu,kuroki,graser,ikeda,
brydon,Luo,kuroki2,kaneshita,bascones,yanagi,Schickling} in general we expect
different 
interaction strengths $U_{\mu,\nu}$ and $J_{\mu,\nu}$ for inequivalent
choices of the orbitals $\mu$, $\nu$. In particular, of the three orbitals
which dominate the electronic structure near  
the Fermi surface, \emph{ab initio} calculations of Miyake \textit{et
al.}\cite{miyake} predict a much weaker interaction on the $xy$ 
orbital compared to the equivalent $xz$ and $yz$ orbitals.
In order to capture this aspect of the physics we renormalize every
interaction strength in Eq.\ (\ref{orb-ind}) involving the $xy$ 
orbital by a multiplicative factor of $V_{xy}\leq 1$; all other interactions
are left unchanged.

\subsection{Method}\label{method}

The static paramagnetic spin susceptibility contains all necessary information
about the magnetic instabilities of the paramagnetic state. In particular, it
diverges at the ordering vector as one approaches the critical temperature
$T_c$. Hence, a temperature-vs.-doping phase diagram for the boundaries of the
paramagnetic phase can be obtained by determining the highest temperature for
which $1/\chi_s({\bf q},0)$ vanishes as the filling is varied, where the
corresponding ${\bf q}$ is the ordering vector ${\bf Q}$. We obtain the static
susceptibility from the analytic continuation $i\omega_n \rightarrow
\omega + i0^+$ of the total susceptibility in Matsubara-frequency space and
subsequently taking the limit $\omega\to 0$. The total spin susceptibility is
defined as 
\begin{align}
  &\chi_s({\bf q},i\omega_n)\nonumber \\
  &= \frac{1}{N}\sum_{j,j'=x,y,z}\sum_{\nu\mu}\int_0^\beta d\tau
  e^{i\omega_n\tau}\left\langle T_\tau S^j_\nu({\bf q},\tau)S^{j'}_\mu(-{\bf
    q},0) \right\rangle. 
\end{align}
The Fourier-transformed spin operator $S^j_\nu({\bf q})$ is related to the
spin operator at site ${\bf i}$ by 
\begin{align}
  S^j_{{\bf i},\nu}=\frac{1}{\sqrt{N}}\sum_{\bf q} S^j_\nu({\bf q})e^{-i{\bf
      q}\cdot{\bf r_i}}.\nonumber 
\end{align}
Due to rotational symmetry in the paramagnetic phase, the spin susceptibility
can be expressed in terms of its transverse part 
\begin{align}
  \chi_s({\bf q},0)=\frac{3}{2}\chi^{-+}({\bf q},0).
\end{align}
The transverse susceptibility is written as
\beq
\chi^{-+}({\bf q},i\omega_n) = \sum_{\nu,\mu}\chi^{-+}_{\nu\nu\mu\mu},
\label{tot_trans}
\eeq
where we introduce the generalized susceptibility
\beqarray
\lefteqn{\chi^{-+}_{\nu\nu'\mu\mu'}({\bf q},i\omega_n)=\frac{1}{N}\sum_{\bf
    k,k'}\int_0^\beta d\tau e^{i\omega_n\tau}} \nonumber \\ 
&& {}\times\left\langle T_\tau d_{{\bf k+q},\nu,\downarrow}^\dagger(\tau)
  d_{{\bf k},\nu',\uparrow}(\tau)d^\dagger_{{\bf k'-q},\mu,\uparrow}(0)d_{{\bf
      k'},\mu',\downarrow}(0) \right\rangle. \label{sus_trans}
\eeqarray
The orbitally resolved susceptibilities $\chi^{-+}_{\nu\nu\mu\mu} \equiv
\chi^{-+}_{\nu\mu}$ are of particular interest, as they contain information
about the contribution of the different orbitals to the instability. 

We calculate $\chi^{-+}_{\nu\mu}$ using the RPA. 
Summing up the ladder diagrams, we obtain the Dyson equation 
\begin{align}
  \chi^{-+}_{\nu\nu\mu\mu}=\chi^{-+(0)}_{\nu\nu\mu\mu}+
  \chi^{-+(0)}_{\nu\nu\alpha\beta}V_{\alpha\beta\gamma\delta}
  \chi^{-+}_{\gamma\delta\mu\mu},
\end{align}
where the non-zero elements of $V_{\alpha\beta\gamma\delta}$ are given by
\begin{subequations}
\beqarray
V_{aaaa} & = & U[1-(1-V_{xy})\delta_{a,xy}] \\
V_{aabb} & = & J[1-(1-V_{xy})(\delta_{a,xy}+\delta_{b,xy})] \\
V_{abba} & = & (U-2J)[1-(1-V_{xy})(\delta_{a,xy}+\delta_{b,xy})] \\
V_{abab} & = & J[1-(1-V_{xy})(\delta_{a,xy}+\delta_{b,xy})] 
\eeqarray
\end{subequations}
with $a\neq b$. The bare susceptibilities $\chi^{-+(0)}_{\nu\nu'\mu\mu'}$ are
given by 
\beqarray
\lefteqn{ \chi^{-+(0)}_{\nu\nu'\mu\mu'}({\bf q},i\omega_n) } \nonumber\\
&= &  -\frac{1}{N}\sum_{\bf k}\sum_{s,s'}u_{s,\nu'}({\bf k})u^*_{s,\mu}({\bf
    k})u_{s',\mu'}({\bf k+q})u^*_{s',\nu}({\bf k+q}) \nonumber \\
&& {}\times\frac{n_F(E_{s,{\bf k}})-n_F(E_{s',{\bf k+q}})}{E_{s,{\bf
        k}}-E_{s',{\bf k+q}}-i\omega_n}, 
\label{chiNull}
\eeqarray
where $n_F(E)$ is the Fermi function, $E_{{\bf k}}$ are the eigenvalues
of $H_0$, and $u_{s,\nu}({\bf k})$ are the coefficients that transform
the annihilation operators of the diagonalizing basis $\gamma_{s,{\bf k}}$
into the orbital basis, i.e., $d_{\nu,{\bf k}}=\sum_s u_{s,\nu}({\bf
k})\gamma_{s,{\bf k}}$. We adopt the common approximation to ignore Hartree
shifts,\cite{kuroki2,graser,kaneshita,Luo} as we assume them to be included in
the {\it ab initio} calculations. We have verified that the inclusion of Hartree
shifts only leads to small quantitative changes in our results.

\section{Results}\label{results}

\begin{figure}[t]
  \begin{center}
    \includegraphics[clip,width=\columnwidth]{figure1.eps}
    \caption{(color online) Band structure along high-symmetry lines
      for the four models of Refs.\ \onlinecite{kuroki,graser,ikeda,calderon}.} 
    \label{compareBands}
  \end{center}
\end{figure}

\begin{figure}[t]
  \begin{center}
    \includegraphics[clip,width=\columnwidth]{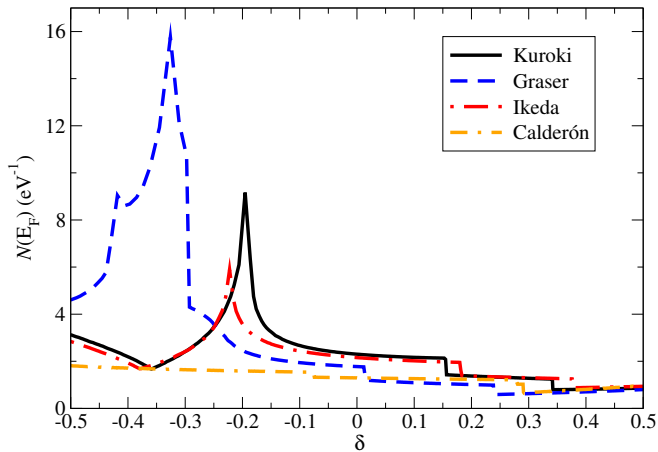}
    \caption{(color online) Density of states at the Fermi level
      as a function of the doping $\delta$, related by the electron
      concentration by $n=6+\delta$, for the four band structures from
      Fig.\ \ref{compareBands}.} 
    \label{compareDOS}
  \end{center}
\end{figure}

\begin{figure}[t]
  \begin{center}
    \includegraphics[clip,width=\columnwidth]{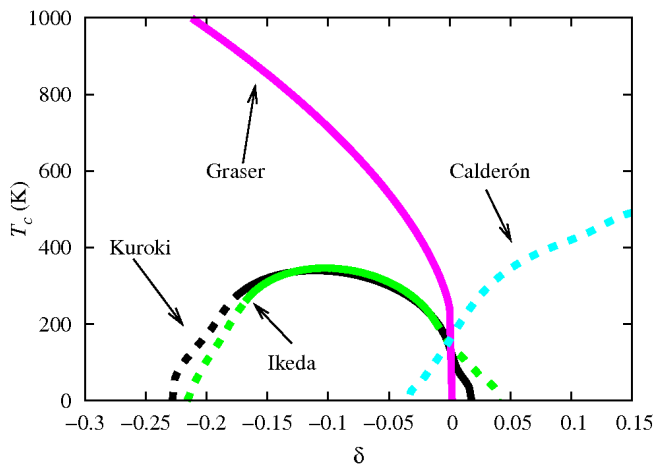}
    \caption{(color online) Critical temperature as a function of doping 
      $\delta$ for the four models from Fig.\ \ref{compareBands} with
      $V_{xy}=1$, see text. Solid lines
      denote $(\pi,0)$ order, whereas dashed lines denote incommensurate
      order. The interaction strengths were chosen as $U=0.875$ eV for
      Kuroki \textit{et al.};\cite{kuroki} $U=0.885$ eV for Ikeda \textit{et
      al.},\cite{ikeda}
      $U=1.2232$ eV for Graser \textit{et al.},\cite{graser} and $U=1.383$ eV
      for Calder\'{o}n \textit{et al.}\cite{calderon}} 
    \label{phase_fin0}
  \end{center}
\end{figure}

In this section we present our analysis of the four different five-orbital
models.\cite{kuroki,graser,ikeda,calderon}
We first investigate the temperature-vs.-doping phase
diagrams, calculated as described in section \ref{method}. The summation
over ${\bf k}$ in Eq.\ (\ref{chiNull}) was performed using a $400 \times 400$
${\bf k}$-point mesh. We have checked that a $200 \times 200$ ${\bf k}$-point
mesh results in only small quantitative differences and so finite size effects
are expected to be negligible. We ignore ordered states with critical
temperatures less than $10$K. Throughout we measure the doping $\delta$
relative to parent compound filling, i.e., the electron concentration is
$n=6+\delta$.

We start by discussing the non-interacting system. In Fig.\ \ref{compareBands}
we plot the non-interacting band structure of the undoped models close
to the Fermi energy. The models of Kuroki
\textit{et al.}\cite{kuroki} and Ikeda \textit{et al.}\cite{ikeda} are hardly
distinguishable within about $0.1$~eV of the Fermi surface, and so we expect a
very similar phase diagram for
these two models at weak doping. Near to the $\Gamma$ point, the Fermi
surfaces for the model of Graser \textit{et al.}\cite{graser} almost coincide
with those of Kuroki \textit{et al.}\ and 
Ikeda \textit{et al.}, but elsewhere there are significant differences: The hole
pocket at the M point is very small compared to 
these models, the electron pocket at the X point is
also smaller, and the 
$(3z^2-r^2)$-derived flat band at the M point lies much closer to the Fermi
energy. The
model of Calder\'{o}n \textit{et al.}\cite{calderon} is quite distinct from the
other models, with highly elliptical electron pockets, almost degenerate hole
pockets at the $\Gamma$ point, and no flattening of the
$(3z^2-r^2)$-derived band at the M point.

The density of states as a function of doping, shown in Fig.\ \ref{compareDOS},
provides additional insight. The models of Ikeda \textit{et al.}\cite{ikeda}
and Kuroki
\textit{et al.}\cite{kuroki} both show a peak in the density of states close to
$\delta=-0.2$, although it is significantly larger in the latter. This peak
arises from the flat bottom of the electron band at the X point. The model
of Graser \textit{et al.}\cite{graser} also shows a very high density of states
below $\delta=-0.3$ that is connected to the $(3z^2-r^2)$-derived flat band at
the M point. These peaks raise the possibility of a
competition between AFM and ferromagnetic order at strong hole  
doping.\cite{xu} In contrast, the model of Calder\'{o}n
\textit{et al.}\cite{calderon} has an almost featureless density of states,
due to the much lower bottom of the electron bands and the absence of the
$(3z^2-r^2)$-derived flat band at the M point.

In order to study the doping dependence of the AFM phase we first choose the 
interaction strength $U$ for each model such that at $\delta=0$ the critical
temperature is $T_c\approx150$~K, close to the experimentally observed
ordering temperature. The resulting temperature-vs.-doping phase diagrams for
all four models are shown in Fig.\ \ref{phase_fin0}.
All models show an
enhancement of $T_c$ for non-zero doping. The dome structure of $T_c$
with a maximum at moderate hole doping ($\delta\approx-0.1$)
for the models of Kuroki \textit{et al.}\cite{kuroki} and Ikeda \textit{et
al.}\cite{ikeda} is
consistent the phase diagram determined by Ikeda \textit{et al.}\ in
Ref.\ \onlinecite{ikeda}. In contrast, the monotonous increase of $T_c$ in the 
model of Graser \textit{et al.}\cite{graser} for strong hole doping, and the
enhancement
of magnetic order for {\it electron} doping in the model of Calder\'{o}n
\textit{et al.},\cite{calderon} are contradicted by experiments. The
model of Calder\'{o}n \textit{et al.}\ further deviates from experimental
findings by displaying a highly incommensurate ordering vector ${\bf
Q}=(\pi,0.24\pi)$ at zero doping.  

The models of Kuroki \textit{et al.}\ and Ikeda \textit{et al.}\ are most
consistent with the
reported asymmetric doping-dependence of the AFM in the pnictides,
although the optimal critical temperature is much too high. Note that there is
no reason to assume that the
undoped \emph{three}-dimensional parent compounds are best described by
choosing precisely the value $\delta=0$ in these \emph{two}-dimensional
models.
Moreover, we will see that a moderate increase of the onsite energy of the $xy$
orbital shifts the peak position to $\delta=0$ for some of the models.
Instead of specifying $T_{c}$ at $\delta=0$ as in Fig.\ \ref{phase_fin0},
it is therefore more
reasonable to search for an interaction strength giving a dome shape of
$T_c$ and an ordering temperature of $T_c^{\rm opt}\approx165$~K at a non-zero
optimal doping $\delta$. The lower $T_{c}^{\rm opt}$ should also
allow us to unambiguously identify the leading nesting instabilities
responsible for the SDW. Below we construct the phase diagram of each model
according to this argument.

\subsection{The model of Kuroki \textit{et al.}}\label{kuroki}

\begin{figure}[t]
  \begin{center}
    \includegraphics[clip,width=\columnwidth]{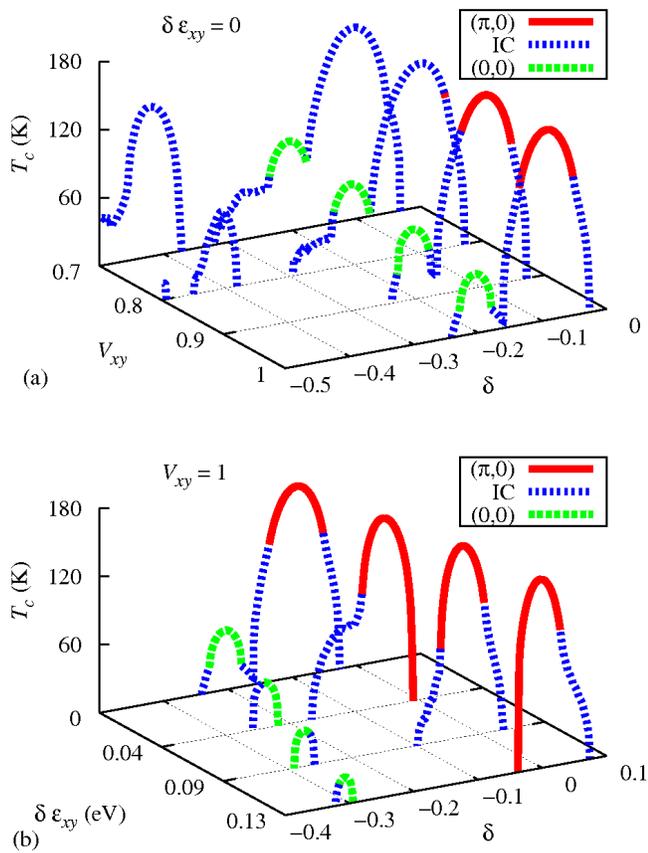}
    \caption{(color online)  (a) Doping dependence of the critical
      temperature for the model of Kuroki \textit{et al.}\cite{kuroki} at
      different values of $V_{xy}$. To maintain $T_c^{\rm
        opt}\approx165$~K, we choose $U=0.8$ eV for $V_{xy}=1$, $U=0.845$ eV for
      $V_{xy}=0.9$, $U=0.885$ eV for $V_{xy}=0.8$, and $U=0.915$ eV for
      $V_{xy}=0.7$. (b) Doping dependence
      of the critical temperature for different shifts of the $xy$-orbital
      on-site 
      energy $\delta\varepsilon_{xy}$ with respect to the
      value given in Ref.\ \onlinecite{kuroki}. For constant optimal doping
      critical temperature 
      $T_c^{\rm opt}\approx165$ K we choose $U=0.8$ eV for
      $\delta\varepsilon_{xy}=0$ eV,
      $U=0.782$ eV for $\delta\varepsilon_{xy}=1.04$ eV, $U=0.78$ eV for
      $\delta\varepsilon_{xy}=1.09$ eV, and $U=0.785$ eV for
      $\delta\varepsilon_{xy}=1.13$ eV.} 
    \label{kuroki-3dphase}
  \end{center}
\end{figure}

\begin{figure}[t]
  \begin{center}
    \includegraphics[clip,width=\columnwidth]{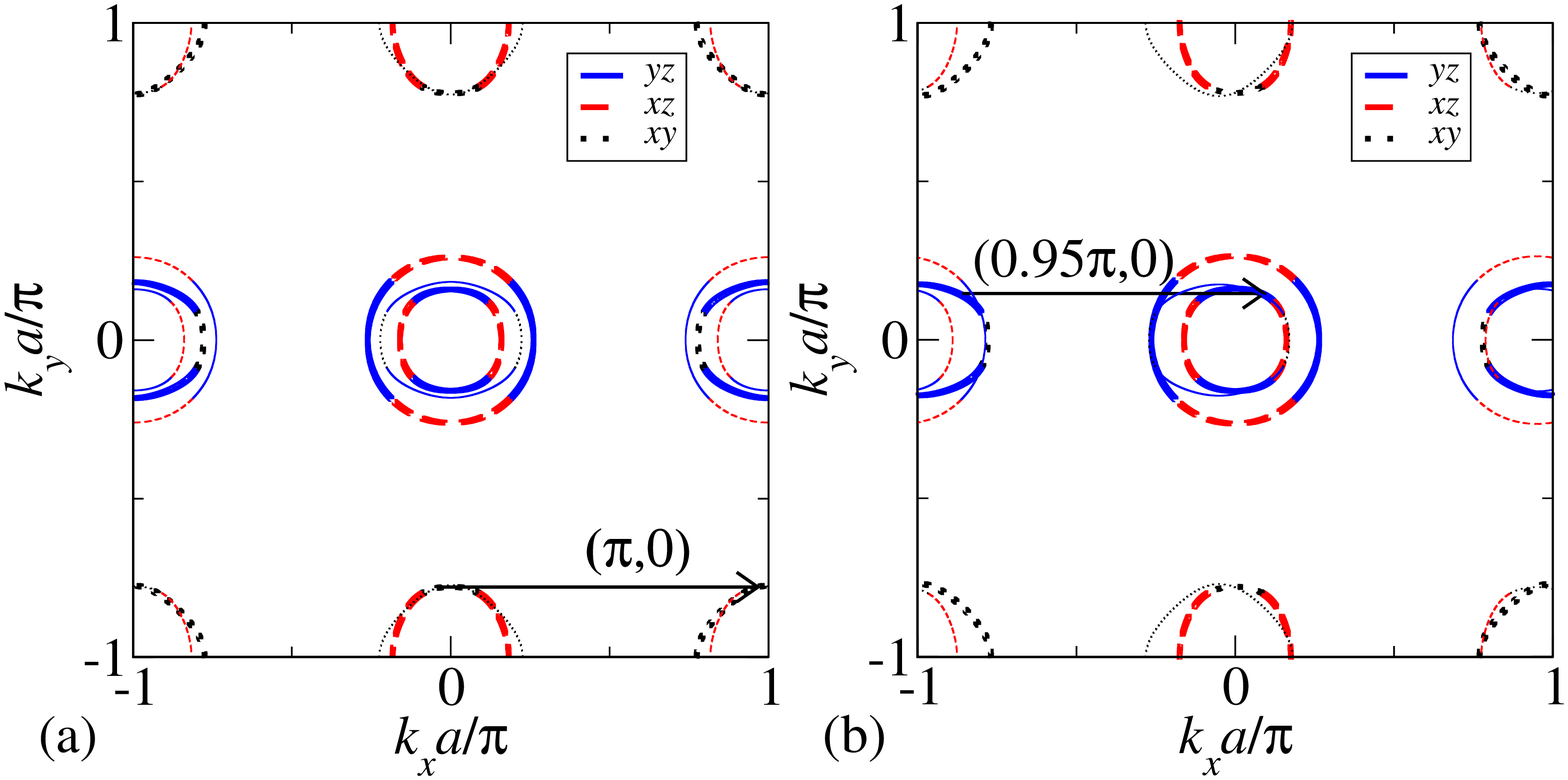} 
    \caption{(color online) Fermi surface for the model of Kuroki
      \textit{et al.}\cite{kuroki}  (heavy lines) superimposed with the Fermi
surface      shifted by (a) $(\pi,0)$ for $\delta=-0.093$ and (b) $(0.95\pi,0)$
      for $\delta=-0.105$ (thin lines), corresponding to the ordering at optimal
      doping for $V_{xy}=1$ 
      and $V_{xy}=0.7$, respectively.} 
    \label{kuroki-FS}
  \end{center}
\end{figure}

\begin{figure}[t]
  \begin{center}
    \includegraphics[clip,width=\columnwidth]{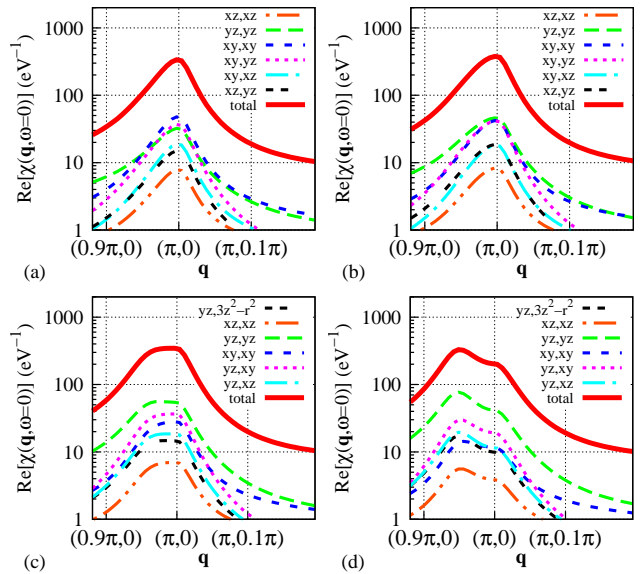}
    \caption{(color online) Total susceptibilities and orbitally
      resolved contributions $\chi^{-+}_{\nu\mu}$ at $T=180$ K for the
      model of Kuroki {\it et al.}\cite{kuroki} at (a)
      $V_{xy}=1$, $\delta=-0.093$, (b) $V_{xy}=0.9$, $\delta=-0.092$, (c)
      $V_{xy}=0.8$, $\delta=-0.096$, and (d) $V_{xy}=0.7$, $\delta=-0.105$.} 
    \label{kuroki-susc}
  \end{center}
\end{figure}

\begin{figure}[t]
  \begin{center}
    \includegraphics[clip,width=\columnwidth]{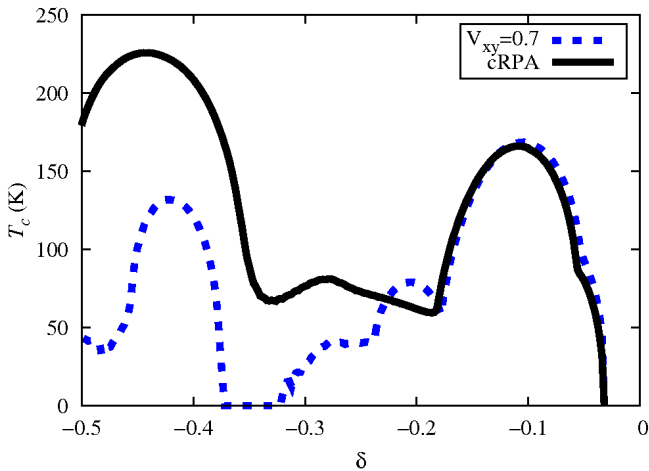}
    \caption{(color online)  Doping dependence of the critical temperature
      for the model of Kuroki \textit{et al.}\cite{kuroki} for $V_{xy}=0.7$
      (dashed) and for  the fully orbital-dependent interactions
      predicted in Ref.\ \onlinecite{miyake}
      (solid).  For the latter, the interactions strengths were
      scaled by $0.406$ to find a maximal ordering
      temperature of $T_c\approx165$ K at $\delta>-0.2$. Both calculations
      yield an ordering 
      vector ${\bf Q}=(0.95\pi,0)$ at the maximum $T_c$ close to
      $\delta=-0.1$.}
    \label{phase_orbres}
  \end{center}
\end{figure}

The phase diagram for the model of Kuroki \textit{et al.}\cite{kuroki}  is
shown in Fig.\ \ref{kuroki-3dphase}. Initially focusing on the
case of $V_{xy}=1$, we still find a dome of $(\pi,0)$ order centered at the
optimal doping $\delta=-0.093$, but there is no magnetic order at $\delta=0$. In
addition to the AFM dome, there is also a region of
ferromagnetic order around $\delta=-0.2$. The highest critical temperature of
the ferromagnetic state occurs close to the peak in the density of
states, see Fig.\ \ref{compareDOS}, consistent with the Stoner criterion.  

The mechanism responsible for the $(\pi,0)$ order can be
observed most clearly at  the optimal doping $\delta=-0.093$. As shown in
Fig.\ \ref{kuroki-FS}(a), here we find excellent nesting of the
$xy$-orbital-dominated parts of the electron 
pocket at the Y point with the $xy$-derived hole pocket at the M point,
suggesting that these Fermi surfaces play
the leading role in the AFM instability. A further hint
that the $xy$ orbital is most important for the AFM order comes
from the observation that the optimal doping can be shifted to $\delta=0$ by
increasing the on-site energy of the $xy$ orbital by $\sim0.1$ eV 
[see Fig.\ \ref{kuroki-3dphase}(b)],
and always coincides with good nesting of the Y and M pockets. 

To examine the effect of the likely weaker interactions in the $xy$
orbital, in Fig.\ \ref{kuroki-3dphase}(a) we plot the 
evolution of the phase diagram upon reducing $V_{xy}$ while simultaneously
increasing $U$ such that $T_c^{\rm opt}$ remains constant. At $V_{xy}=0.7$,
which  is close to the value predicted 
by Miyake \textit{et al.},\cite{miyake} we find that the optimal doping of the
AFM dome shifts to $\delta=-0.105$, and the ordering vector at this filling
becomes weakly incommensurate with ${\bf Q}=(0.95\pi,0)$. The Fermi surface
plotted in
Fig.\ \ref{kuroki-FS}(b) shows that this corresponds to good nesting
between the hole pockets at the $\Gamma$ 
point and the electron pocket at the X point, where the best-nested
segments of the Fermi surface have mostly $yz$-orbital character. The
necessary increase of $U$ by $\sim10\%$ when $V_{xy}$ is reduced
implies that the AFM instability due to $\Gamma$-X nesting requires a
significantly higher interaction than the Y-M nesting to produce a realistic
$T_c^{\rm opt}$. Additional 
magnetically ordered states appear at strong doping $\delta\approx-0.42$ with
optimal ordering vector ${\bf Q}=(0.44\pi,0)$. At these doping levels,
however, our assumption of rigid bands is questionable and so the physical
relevance of these results is doubtful. 

Greater insight into the origin of the AFM order can be achieved by
examining the orbitally resolved susceptibilities at optimal doping just above
$T_c$, see Fig.\ \ref{kuroki-susc}. As $V_{xy}$ is decreased, the dominant 
contribution shifts from the  susceptibilities
$\chi^{-+}_{\nu\mu}$ involving the $xy$ orbital to those involving the $yz$
orbital, and the peak in
$\chi^{-+}_{\nu\mu}$ moves from $(\pi,0)$ to an incommensurate vector. 
This is in perfect agreement with the observed change of the nesting from
$xy$-dominated to $yz$-dominated parts of the Fermi surface. By changing
$V_{xy}$ we can therefore select the dominant nesting instability of
the system.

Our treatment of the weaker interaction on the $xy$ orbital 
neglects the likely different interaction strengths involving the other
inequivalent orbitals. To test our approximation, we compare the
$V_{xy}=0.7$ phase diagram of Fig.\ \ref{kuroki-3dphase}(a) with the phase
diagram calculated using Miyake \emph{et al.}'s\cite{miyake} cRPA values for
$U_{\nu\mu}$ and
$J_{\nu\mu}$ in Eq.\ (\ref{orb-dep}). As shown in Fig.\ \ref{phase_orbres}, upon
suitably
rescaling the cRPA interaction potentials, we find excellent agreement between
the two phase diagrams for the physically reasonable doping regime
$\delta>-0.2$. We note that in the fully 
orbital-dependent results we have to choose a slightly larger value of
$U_{yz,yz}$ than in our $V_{xy}=0.7$ calculations in order to achieve
$T_c^\mathrm{opt} \approx 165$ K for the peak around $\delta\approx -0.1$; the
origin of this discrepancy is likely the weaker exchange interaction
$J_{\nu\mu}<0.25\,U_{\nu\nu}$ predicted by Ref.\ \onlinecite{miyake}. This also
indicates that the AFM order does not crucially depend upon the ratio $U/J$
in the weak-coupling regime. For strong 
doping, larger deviations appear, in particular 
there is no ferromagnetism for the cRPA interactions, and the critical
temperature of the incommensurate AFM state at $\delta\approx-0.4$ is higher
although the ordering vector is similar.

\subsection{The model of Graser \textit{et al.}}

\begin{figure}[t]
  \begin{center}
    \includegraphics[clip,width=\columnwidth]{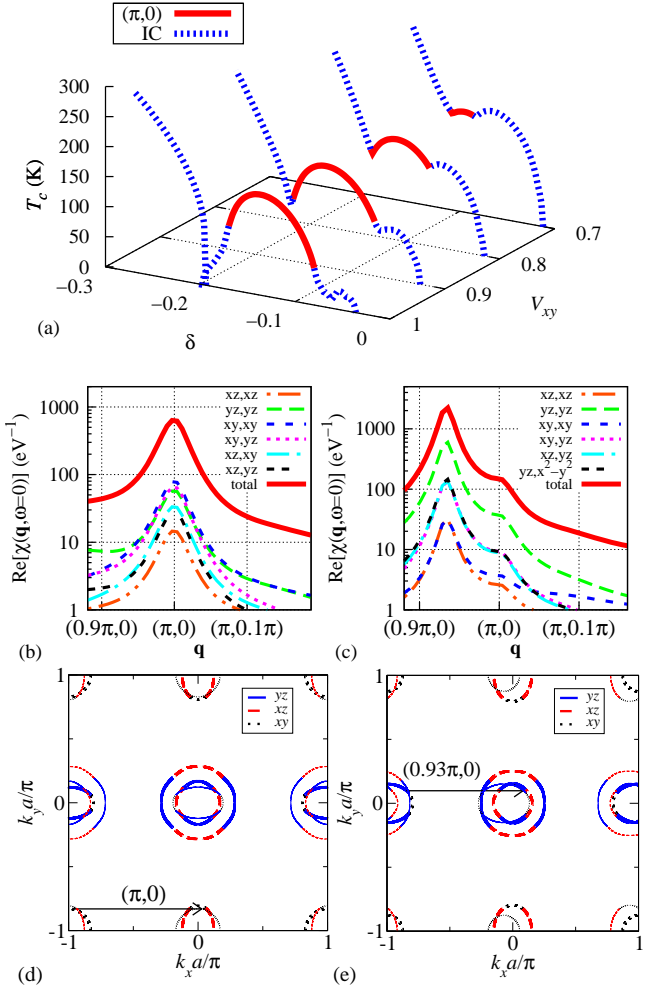}
    \caption{(color online)  (a) Doping dependence of the critical
      temperature for the model of Graser
      \textit{et al.}~\cite{graser} at
      different values of $V_{xy}$. To maintain $T_c^{\rm
        opt}\approx165$~K, we choose $U=1.09$~eV for 
      $V_{xy}=1$, $U=1.146$~eV for $V_{xy}=0.9$, $U=1.191$~eV for
      $V_{xy}=0.8$, and $U=1.222$~eV for $V_{xy}=0.7$. (b) The total
      susceptibility and the largest 
      $\chi^{-+}_{\nu\mu}$ for $V_{xy}=1$, $T=180$~K, and the optimal doping
      $\delta=-0.137$ of the 
      $(\pi,0)$ state. (c) Same as in (b) but for $V_{xy}=0.7$, $T=180$~K, and
      the
      optimal doping $\delta=-0.065$ of the incommensurate state. (d)
      Fermi surface at $\delta=-0.137$ (heavy lines) with the same
      Fermi surface shifted by $(\pi,0)$ superimposed (thin lines). (e)
      Fermi surface at $\delta=-0.065$ (heavy lines) with the same Fermi
      surface shifted by $(0.93\pi,0)$ superimposed (thin lines).} 
    \label{graser_plots}
  \end{center}
\end{figure}

The phase diagram for the model of Graser \textit{et al.}\cite{graser} is shown
in Fig.\ \ref{graser_plots}(a). At $V_{xy}=1$ we find a dome of commensurate
AFM order centered at $\delta=-0.12$ and also a small incommensurate
dome with 
${\bf Q}=(0.95\pi,0)$ at $\delta\approx-0.05$. At 
$\delta<-0.2$ we observe an incommensurate AFM state with ordering vector
${\bf Q}=(0.71\pi,0)$, and a high critical temperature, which strongly
increases with hole doping. The strong tendency to
AFM order at strong hole doping occurs only in this model, and is
likely connected with the ($3z^2-r^2$)-derived flat band at the
M point. Indeed, a Fermi surface due to this
band appears at the critical doping level for the incommensurate AFM
order. Furthermore, the 
ordering at strong doping can be suppressed by increasing the on-site energy
of the $xy$ orbital, which effectively lowers the flat band at the M point.
This also shifts the optimal doping of the $(\pi,0)$ dome at
$\delta=-0.12$ towards zero, again suggesting an important role for the $xy$
orbital in the $(\pi,0)$ order. 

Focusing our attention on the regime of moderate doping, $-0.2<\delta<0$, we
find 
that as $V_{xy}$ is reduced, the small incommensurate dome grows and becomes
the leading instability at $V_{xy}=0.7$. At this value of $V_{xy}$, the
ordering vector at optimal doping $\delta=-0.06$ is ${\bf Q}=(0.93\pi,0)$.
The orbitally resolved susceptibilities at the optimal doping
of the $(\pi,0)$  order at $V_{xy}=1$ and the incommensurate 
order for $V_{xy}=0.7$  are plotted in Figs.~\ref{graser_plots}(b) and
(c), respectively, while the nesting of 
the corresponding Fermi surfaces are shown in Figs.~\ref{graser_plots}(d) and
(e). These results are very similar to those obtained for the model
of Kuroki \textit{et al.},\cite{kuroki} see Figs.~\ref{kuroki-susc} and
\ref{kuroki-FS}. We hence conclude that again a reduction of $V_{xy}$
tunes the system from the $xy$-dominated to the $yz$-dominated instability,
although the optimal dopings for the two instabilities are more widely separated
than in the model of Kuroki 
\textit{et al.} This can be explained by the small M pocket at zero doping in
Graser \textit{et al.}'s model,\cite{graser} which implies a much larger change
of the doping in order to optimize the Y-M nesting. Furthermore, due to the
smaller size of the electron pockets at parent compound filling, the
incommensurate AFM phase is stabilized at weaker doping.

\subsection{The model of Ikeda \textit{et al.}}

\begin{figure}[t]
  \begin{center}
    \includegraphics[clip,width=\columnwidth]{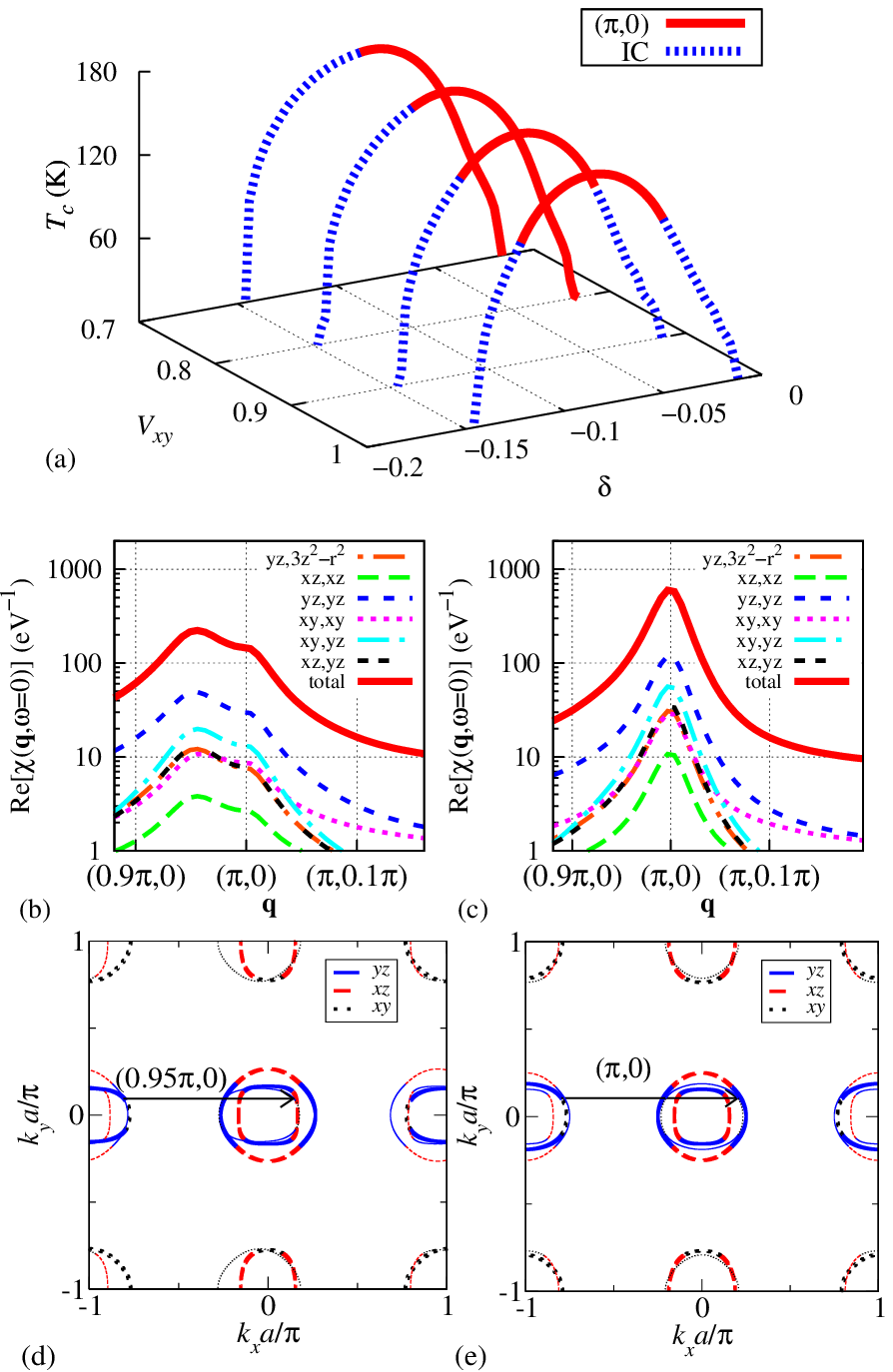}
    \caption{(color online)  (a) Doping dependence of the critical
        temperature for the model of Ikeda 
      \textit{et al.}\cite{ikeda} at
      different values of $V_{xy}$. To maintain $T_c^{\rm
        opt}\approx165$~K, we choose $U=0.808$~eV for
      $V_{xy}=1$, $U=0.852$~eV for $V_{xy}=0.9$, $U=0.892$~eV for
      $V_{xy}=0.8$, and $U=0.925$~eV for $V_{xy}=0.7$.  (b) The total
      susceptibility and the largest 
      $\chi^{-+}_{\nu\mu}$ for $V_{xy}=0.7$, $T=160$~K, and $\delta=-0.12$. (c)
      Same as in (b) but at $\delta=-0.06$. (d)
      Fermi surface at $\delta=-0.12$ (heavy lines) with the same
      Fermi surface shifted by $(0.95\pi,0)$ superimposed (thin lines). (e)
      Fermi surface at $\delta=-0.06$ (heavy lines) with the same Fermi
      surface shifted by $(\pi,0)$ superimposed (thin lines).} 
    \label{ikeda_plots}
  \end{center}
\end{figure}

The  phase diagram for the model of Ikeda
\textit{et al.}\cite{ikeda}  at $V_{xy}=1$ [Fig.\ \ref{ikeda_plots}(a)] is very
similar to that for the model 
of Kuroki \textit{et al.},\cite{kuroki} although ferromagnetism is not
found near $\delta=-0.2$, consistent with the smaller peak in the density of
states [Fig.\ \ref{compareDOS}]. The $(\pi,0)$ order is again  dominated by
the $xy$ orbital, and the nesting of the Y and M pockets is
primarily responsible for the AFM state, in agreement with Ref.\
\onlinecite{ikeda}. Despite the very similar band structure to
the model of Kuroki \textit{et al.}, the phase diagram for $V_{xy}<1$ shows a 
significant difference: The $(\pi,0)$ order does not vanish for $V_{xy}=0.7$,
but instead moves to the low-doping half of the AFM dome while the other half
is incommensurate. As for Kuroki \textit{et al.}'s model, however, the $yz$
orbital dominates the magnetism over 
the full doping range. This can be seen in two representative
plots of $\chi^{-+}_{\nu\mu}$ in Figs.~\ref{ikeda_plots}(b) and (c) at
$\delta=-0.12$ and $\delta=-0.06$ which correspond to incommensurate
${\bf Q}=(0.95\pi,0)$ and commensurate
${\bf Q}=(\pi,0)$ ordering vectors, respectively. As revealed
by Figs.~\ref{ikeda_plots}(d) and (e), at both of these doping levels
there is excellent nesting of the X and $\Gamma$ pockets. The ordering
vector tracks the continuous evolution of this nesting vector across the
dome, from slightly incommensurate to commensurate.
The close similarity of this model to the one of Kuroki
\textit{et al.}\ shows that very
small differences in the band structure play a major role in determining
the ordering vector.

\subsection{The model of Calder\'{o}n \textit{et al.}}

\begin{figure}[t]
  \begin{center}
    \includegraphics[clip,width=\columnwidth]{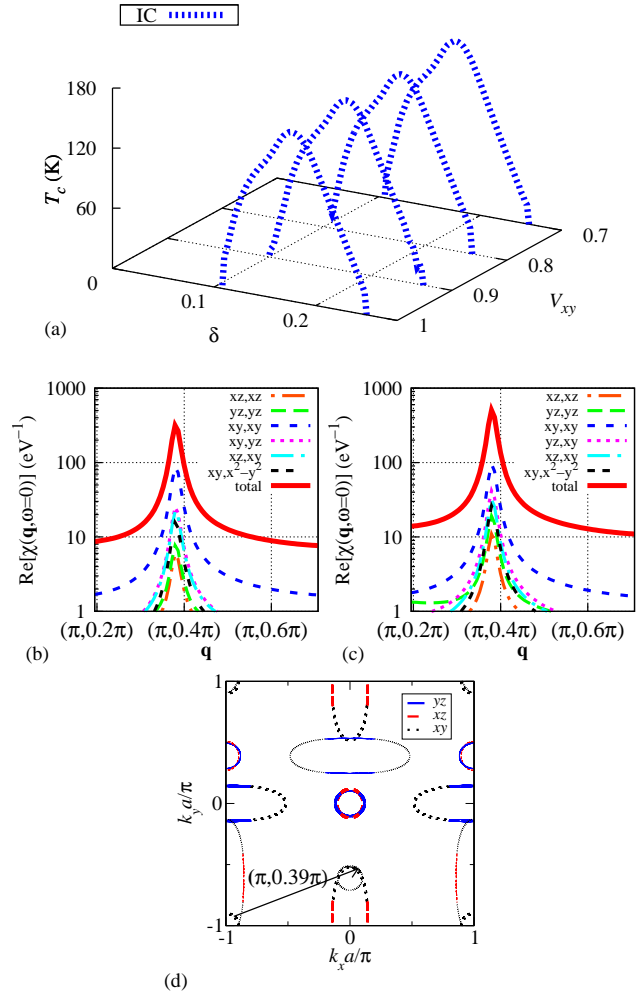}\\
    \caption{(color online)  (a) Doping dependence of the critical
        temperature for the model of Calder\'{o}n 
      \textit{et al.}\cite{calderon} at
      different values of $V_{xy}$. To maintain $T_c^{\rm
        opt}\approx165$~K, we choose $U=1.269$~eV for $V_{xy}=1$, $U=1.378$~eV
      for $V_{xy}=0.9$, $U=1.498$~eV 
      for $V_{xy}=0.8$ and $U=1.635$~eV for $V_{xy}=0.7$.   (b) The total
      susceptibility and the largest 
      $\chi^{-+}_{\nu\mu}$ for $V_{xy}=1$, $T=180$~K, and $\delta=0.176$. (c)
      Same as in (b) but for $V_{xy}=0.7$. (d)
      Fermi surface at $\delta=0.176$ (heavy lines) with the same
      Fermi surface shifted by $(\pi,0.39\pi)$ superimposed (thin lines).} 
    \label{calderon_plots}
  \end{center}
\end{figure}

The phase  diagram for the model of Calder\'{o}n
\textit{et al.},\cite{calderon} shown in Fig.\ \ref{calderon_plots}(a), is in
stark contrast to those for
the other models. The magnetic order is optimized at 
strong electron doping, $\delta\approx 0.18$, and occurs at the highly
incommensurate ordering vector ${\bf Q}=(\pi,0.39\pi)$. Although this
is inconsistent with experimental findings for the pnictide systems,
it is nevertheless interesting to examine the origin of 
this AFM state. An important clue comes from observing that the
phase diagram hardly changes when $V_{xy}$ is decreased, although the
interaction $U$ has to be increased in
order to keep $T_c^{\rm opt}$ constant. This indicates that the $xy$ orbital
is almost exclusively responsible for the magnetic ordering,  which is
confirmed by the orbitally resolved susceptibilities presented in
Figs.~\ref{calderon_plots}(b) and (c). The incommensurate ordering vector
at optimal doping gives excellent nesting of the $xy$ pocket at the M
point and the $xy$-dominated tip of the electron Fermi surface at 
the Y point [Fig.~\ref{calderon_plots}(d)], revealing an unexpected similarity
to the nesting instabilities in the other models. 

The stabilization of AFM order by electron-doping in the model of
Calder\'{o}n \textit{et al.} follows from the observation that the magnetic
order arises only from Y-M nesting. In
Fig.\ \ref{compareBands} it can be seen that at $\delta=0$ the circular hole
pocket at the M point has almost the same radius as in the models of Kuroki
\textit{et al.}\cite{kuroki} and Ikeda \textit{et al.}\cite{ikeda}
In contrast, the ellipticity of the electron pockets is highly exaggerated
in the model of Calder\'{o}n \textit{et al.}, and indeed the minor axis of
the electron pockets 
is much smaller than the diameter of the M-point hole pocket. As such, it is
necessary to raise the chemical potential (i.e., dope with electrons)
to optimize the nesting between these two Fermi surfaces.

\section{Discussion}\label{discussion}

Our study of the four different models for the 1111 pnictides
shows a clear distinction between the models of Kuroki \textit{et
al.},\cite{kuroki}  Graser \textit{et al.},\cite{graser} and Ikeda \textit{et
al.}\cite{ikeda} on the one hand and the
model of Calder\'{o}n \textit{et al.}\cite{calderon} on the other. In the
former, a SDW is stabilized for hole doping with commensurate or
near-commensurate ordering vector, whereas in the latter we find a strongly
incommensurate AFM state upon electron doping. The behavior of Calder\'{o}n
\textit{et al.}'s model can be explained by the unrealistically high ellipticity
of the electron pockets.
Calder\'{o}n \textit{et al.}\cite{calderon} started
from an eight-orbital model with nearest-neighbor Fe-Fe and Fe-As hopping
and then removed the As orbitals within perturbation theory. By using the
Slater-Koster approach, they were able to write the eighteen hopping integrals
as functions of only four overlap integrals and the Fe-As bond angle; in
the other models, these eighteen hopping integrals are free
parameters, which gives much greater freedom in fitting the band structure.
One can speculate that the smaller parameter space available to
Calder\'{o}n \textit{et al.}\ is responsible for the unphysical doping
dependence in their model. Among the remaining models, the
monotonous increase of $T_c$ for the model of Graser
\textit{et al.}, see Figs.~\ref{phase_fin0} and \ref{graser_plots}(a), also
contradicts available experimental
data.~\cite{mu,mu2,mu3,wen,bernhard,chen,rotter2009} The
comparably high ($3z^2-r^2$)-derived flat band at the M point is responsible for
this strong tendency to AFM order.\cite{brydon} 
In contrast, the models of Kuroki \textit{et al.}\ and Ikeda \textit{et al.}\
show a dome of
$(\pi,0)$ order centered at moderate hole doping $\delta\approx-0.1$; for
realistic maximum critical temperature $T_c\approx165$~K there is no ordering
at zero nominal doping, $\delta=0$. Similar behavior is seen in the model of
Graser \textit{et al.}\ for
sufficiently small interaction strength. In the first three models there are
hence strong $(\pi,0)$ spin fluctuations at weak doping, consistent with
previous studies~\cite{graser,kuroki,ikeda} and the asymmetric doping
dependence of the SDW phase observed in some
experiments.\cite{mu,mu2,mu3,wen,bernhard,chen,rotter2009,lumsden,chu}

Our results again demonstrate the sensitive dependence of the magnetic order on
small details of the band structure in the weak-coupling limit.\cite{brydon}
This is illustrated by the
major discrepancies between the model of Graser \textit{et al.}\ and the more
realistic models of Kuroki \textit{et al.}\ and Ikeda \textit{et al.},
  despite the very similar band structures of these three models. It is
significant that the models of Kuroki \textit{et al.}\ and Ikeda \textit{et
al.}\ were obtained by fitting to \textit{ab initio} results for the
experimental crystal structure of the FeAs planes, whereas the model of Graser
\textit{et al.}\ is based upon \textit{ab initio}
calculations for a relaxed structure. These two structures have rather
different Fe-As bond angles, which has been identified as a crucial control
parameter for pnictide physics.~\cite{calderon,kuroki,ikeda}

In addition to the major role of the band structure,
important details of the antiferromagnetic order were found to be controlled by
the interaction strengths involving the $xy$ orbital relative to
the other orbitals. Under the common assumption of orbitally rotation-invariant
interactions, the models of Kuroki \textit{et al.}\cite{kuroki} and Ikeda
\textit{et al.}\cite{ikeda} display $(\pi,0)$ order at optimal doping (maximum
SDW critical temperature $T_c$). Upon reducing the interactions involving the
$xy$ orbital,
the optimal doping slightly shifts and the ordering vector tends to become
weakly incommensurate: At the lowest reduction factor $V_{xy}=0.7$, the AFM
dome for the model of Kuroki \textit{et al.}\ becomes entirely incommensurate,
while for the model of Ikeda \textit{et al.}\ there is a continuous change from
incommensurate to
commensurate order as one moves from stronger to weaker doping.
Reducing $V_{xy}$ in the model of Graser
\textit{et al.},\cite{graser} we find that the dome with $(\pi,0)$ order is
almost obscured by a second dome with weakly incommensurate order that has a
smaller optimal doping.

We find that in all models commensurate $(\pi,0)$ order is
realized for an extended doping range. This shows that commensurate
order can be stabilized in purely electronic models, but does
of course not imply that mechanisms beyond the models considered here are
unimportant in this respect. Indeed, the experimentally observed robustness of
commensurate order suggests that some additional stabilizing mechanism is
required, especially in the scenario of reduced $V_{xy}$.
Accounting for the three-dimensionality of the Fermi surface
may improve the nesting with a commensurate ordering vector. Most likely,
though, the weak incommensuration we have found will result in
commensurate order if magneto-elastic coupling is taken into account.

Our results allow us to distinguish the dominant
nesting instability in the various models. The $(\pi,0)$ order observed for
orbitally rotation-invariant interactions originates mainly from the good 
nesting between the $xy$-derived parts of the Y pocket and the M
pocket, as identified in
Refs.~\onlinecite{ikeda2008,arita,kuroki2,ikeda,ikeda2010}. 
For reduced interactions involving the $xy$ orbital, however, the
nesting instability between the $\Gamma$ and X pockets 
gives the highest ordering temperature, which corresponds to the scenario
in Refs.~\onlinecite{vorontsov,cvetkovic,eremin,yu,brydonEI,knolle2010,
QPinterference,brydon,daghofer,fernandes}.
This mechanism is dominated by the $yz$ orbital and produces a SDW with
the ordering vector ${\bf Q}=(Q_x,0)$, where $0.9\pi<Q_x\leq\pi$. 

Our results therefore confirm the two proposals in the literature
for a nesting instability in the pnictides, with their relative importance
tuned by the parameter $V_{xy}$. This implies a crucial role for 
the As $4p$ orbitals, as their hybridization with the Fe $3d$ orbitals is
ultimately responsible for the strong orbital dependence of
the interaction potentials in an effective $3d$ theory.~\cite{miyake} It is
therefore somewhat unsatisfying that the As orbitals do not enter the
calculations more directly. Indeed, keeping the As $4p$ orbitals reduces the 
variation in the size of the Fe $3d$ Wannier functions, and hence the 
interaction strength is likely to show much less pronounced orbital
dependence.~\cite{vildosola,miyake} A detailed comparison of the spin
fluctuations in a realistic $3d$-$4p$ model with those in a pure $3d$ model
is therefore desirable.~\cite{Schickling}

\section{Summary}\label{summary}

In this work we have presented an analysis of the instabilities responsible 
for magnetic order in the 1111 pnictides. Using the RPA we have determined the
doping dependence of the SDW critical temperature and ordering
vector in four different
five-orbital models. For the three models proposed by Kuroki \textit{et
al.},\cite{kuroki} Graser \textit{et al.},\cite{graser} and Ikeda \textit{et
al.},\cite{ikeda} we find that the observed $(\pi,0)$ magnetic order
is stabilized for hole doping, while in the model of Calder\'{o}n \textit{et
al.}\cite{calderon} an incommensurate SDW phase appears at electron doping,
contradicting experimental results. 
We have studied the relative
importance of the two known nesting instabilities across the phase diagram
of these models, and have identified the relative
interaction strength in the $xy$ orbital as a parameter that tunes the
dominant mechanism leading to magnetic order. We have identified two models as
giving particularly good agreement with experiment, and discussed the band
structure features which lead to the poorer agreement for the others.

\section*{Acknowledgments}

The authors thank M. J. Calder\'{o}n, M. Daghofer, T. Dellmann, P. Materne,
J. Spehling, R. Valent\'{\i},
M. Voj\-ta, and B. Zocher for useful discussions. We acknowledge funding by the
Deutsche Forschungsgemeinschaft through Priority Programme 1458.


\begin{thebibliography}{99}
\bibitem{kamihara} Y. Kamihara, T. Watanabe, M. Hirano, and Hideo Hosono,
  J. Am. Chem. Soc. {\bf 130}, 3296 (2008). 

\bibitem{rotter} M. Rotter, M. Tegel, and D. Johrendt, Phys. Rev. Lett. {\bf
  101}, 107006 (2008). 

\bibitem{zhao} J. Zhao, Q. Huang, C. de la Cruz, S. Li, J. W. Lynn, Y. Chen,
  M. A. Green, G. F. Chen, G. Li, Z. Li, J. L. Luo, N. L. Wang, P. Dai, Nature
  Materials {\bf 7}, 953 (2008); J. Zhao, Q. Huang, C. de la Cruz, J. W. Lynn,
  M. D. Lumsden, Z. A. Ren, J. Yang, X. Shen, X. Dong, Z. Zhao, P. Dai,
  Phys. Rev. B {\bf 78}, 132504 (2008). 

\bibitem{huang} Q. Huang, Y. Qiu, W. Bao, J. W. Lynn, M. A. Green, Y. Chen,
  T. Wu, G. Wu, and X. H. Chen, Phys. Rev. Lett. {\bf 101}, 257003 (2008);
  A. Jesche, N. Caroca-Canales, H. Rosner, H. Borrmann, A. Ormeci,
  D. Kasinathan, K. Kaneko, H. H. Klauss, H. Luetkens, R. Khasanov, A. Amato,
  A. Hoser, C. Krellner, and C. Geibel, Phys. Rev. B {\bf 78}, 180504(R)
  (2008). 

\bibitem{lumsden} M. D. Lumsden and A. D. Christianson, J. Phys.:
  Condens. Matter {\bf 22}, 203203 (2010). 

\bibitem{sebastian} S. E. Sebastian, J. Gillett, N. Harrison, P. H. C. Lau,
  C. H. Mielke, and G. G. Lonzarich, J. Phys.: Condens. Matter {\bf 20}, 422203
  (2008); J. G. Analytis, R. D. McDonald, J.-H. Chu, S. C. Riggs,
  A. F. Bangura, C. Kucharczyk, M. Johannes, and I. R. Fisher, Phys. Rev. B
  {\bf 80}, 064507 (2009). 

\bibitem{yi} M. Yi, D. H. Lu, J. G. Analytis, J.-H. Chu, S.-K. Mo, R.-H. He,
  M. Hashimoto, R. G. Moore, I. I. Mazin, D. J. Singh, Z. Hussain,
  I. R. Fisher, and Z.-X. Shen, Phys. Rev. B {\bf 80}, 174510 (2009).  

\bibitem{shimojima} T. Shimojima, K. Ishizaka, Y. Ishida, N. Katayama,
  K. Ohgushi, T. Kiss, M. Okawa, T. Togashi, X.-Y. Wang, C.-T. Chen,
  S. Watanabe, R. Kadota, T. Oguchi, A. Chainani, and S. Shin,
  Phys. Rev. Lett. {\bf 104}, 057002 (2010). 

\bibitem{mcguire} M. A. McGuire, A. D. Christianson, A. S. Sefat, B. C. Sales,
  M. D. Lumsden, R. Jin, E. A. Payzant, D. Mandrus, Y. Luan, V. Keppens,
  V. Varadarajan, J. W. Brill, R. P. Hermann, M. T. Sougrati, F. Grandjean,
  G. J. Long, Phys. Rev. B {\bf 78}, 094517 (2008); M. A. McGuire,
  R. P. Hermann, A. S. Sefat, B. C. Sales, R. Jin, D. Mandrus, F. Grandjean,
  and G. J. Long, New J. Phys. {\bf 11}, 025011 (2009). 

\bibitem{liu} R. H. Liu, G. Wu, T. Wu, D. F. Fang, H. Chen, S. Y. Li, K. Liu,
  Y. L. Xie, X. F. Wang, R. L. Yang, L. Ding, C. He, D. L. Feng, and
  X. H. Chen, Phys. Rev. Lett. {\bf 101}, 087001 (2008). 

\bibitem{dong} J. K. Dong, L. Ding, H. Wang, X. F. Wang, T. Wu, G. Wu,
  X. H. Chen, and S. Y. Li, New J. Phys. {\bf 10}, 123031 (2008). 

\bibitem{drechsler} S.-L. Drechsler, H. Rosner, M. Grobosch, G. Behr, F. Roth,
  G. Fuchs, K. Koepernik, R. Schuster, J. Malek, S. Elgazzar, M. Rotter,
  D. Johrendt, H.-H. Klauss, B. B\"uchner, and M. Knupfer, arXiv:0904.0827. 

\bibitem{yang} W. L. Yang, P. O. Velasco, J. D. Denlinger, A. P. Sorini,
  C-C. Chen, B. Moritz, W.-S. Lee, F. Vernay, B. Delley, J.-H. Chu,
  J. G. Analytis, I. R. Fisher, Z. A. Ren, J. Yang, W. Lu, Z. X. Zhao, J. van
  den Brink, Z. Hussain, Z.-X. Shen, and T. P. Devereaux, Phys. Rev. B {\bf
    80}, 014508 (2009). 

\bibitem{singh} D. J. Singh and M.-H. Du, Phys. Rev. Lett. {\bf 100}, 237003
  (2008).

\bibitem{mazin} I. I. Mazin, D. J. Singh, M. D. Johannes, and M. H. Du,
  Phys. Rev. Lett. {\bf 101}, 057003 (2008). 

\bibitem{zhang} Y.-Z. Zhang, I. Opahle, H. O. Jeschke, and R. Valent\'{\i},
  Phys. Rev. B {\bf 81}, 094505 (2010). 

\bibitem{wen} H.-H. Wen, G. Mu, L. Fang, H. Yang, and Z. Zhu,
  Europhys. Lett. {\bf 82}, 17009 (2008).

\bibitem{mu3} G. Mu, L. Fang, H. Yang, X. Zhu, P. Cheng, and H.-H. Wen,
  J. Phys. Soc. Jpn. Suppl. {\bf 77}, 15 (2008). 

\bibitem{mu} G. Mu, B. Zeng, X. Zhu, F. Han, P. Cheng, B. Shen, and H.-H. Wen,
  Phys. Rev. B {\bf 79}, 104501 (2009). 

\bibitem{mu2} G. Mu, B. Zeng, P. Cheng, X. Zhu, F. Han, B. Shen, and
  H.-H. Wen, Europhys. Lett. {\bf 89}, 27002 (2010).  

\bibitem{bernhard} C. Bernhard, A. J. Drew, L. Schulz, V. K. Malik, M. R\"ossle,
  C. Niedermayer, T. Wolf, G. D. Varma, G. Mu, H.-H. Wen, H. Liu, G. Wu, and
  X. H. Chen, New J. Phys. {\bf 11}, 055050 (2009). 

\bibitem{chu} J.-H. Chu, J. G. Analytis, C. Kucharczyk, and I. R. Fisher,
  Phys. Rev. B {\bf 79}, 014506  (2009). 

\bibitem{chen} H. Chen, Y. Ren, Y. Qiu, Wei Bao, R. H. Liu, G. Wu, T. Wu,
  Y. L. Xie, X. F. Wang, Q. Huang, and X. H. Chen, Europhys. Lett. {\bf 85},
  17006 (2009). 

\bibitem{rotter2009} M. Rotter, M. Tegel, I. Schellenberg,
  F. M. Schappacher, 
  R. P\"{o}ttgen, J. Deisenhofer, A G\"{u}nther, F. Schrettle,
  A. Loidl, and D. Johrendt, New J. Phys. {\bf 11}, 025014 (2009).

\bibitem{cvetkovic} V. Cvetkovic and Z. Tesanovic, Europhys. Lett. {\bf
    85}, 37002 (2009); Phys. Rev. B {\bf 80}, 024512 (2009).

\bibitem{yu} R. Yu, K. T. Trinh, A. Moreo, M. Daghofer, J. A. Riera, S. Haas,
  and E. Dagotto, Phys. Rev. B {\bf 79}, 104510 (2009). 

\bibitem{brydonEI} P. M. R. Brydon and C. Timm, Phys. Rev. B {\bf 79},
  180504(R) (2009);  Phys. Rev. B {\bf 80}, 174401 (2009)

\bibitem{knolle2010} J. Knolle, I. Eremin, A. V. Chubukov, and R. Moessner,
  Phys. Rev. B {\bf 81}, 140506(R) (2010).

\bibitem{QPinterference} J. Knolle, I. Eremin, A. Akbari, and R. Moessner,
  Phys. Rev. Lett. {\bf 104}, 257001 (2010); A. Akbari, J. Knolle, I. Eremin,
  and R. Moessner, Phys. Rev. B {\bf 82}, 224506 (2010).

\bibitem{daghofer}M. Daghofer, A. Nicholson, A. Moreo, and E. Dagotto,
  Phys. Rev. B {\bf 81}, 014511 (2010). 

\bibitem{eremin}I. Eremin and A. V. Chubukov, Phys. Rev. B {\bf  81}, 024511
  (2010). 

\bibitem{vorontsov} A. B. Vorontsov, M. G. Vavilov, and A. V. Chubukov,
  Phys. Rev. B {\bf 81}, 174538 (2010).

\bibitem{fernandes} R. M. Fernandes and J. Schmalian, Phys. Rev. B {\bf
    82}, 014521 (2010).

\bibitem{brydon} P. M. R. Brydon, M. Daghofer, and C Timm,
  J. Phys.: Condens. Matter {\bf 23}, 246001 (2011). 

\bibitem{ikeda2008} H. Ikeda, J. Phys. Soc. Japan {\bf 77}, 123707
  (2008).

\bibitem{kuroki2} K. Kuroki, H. Usui, S. Onari, R. Arita, and H. Aoki,
  Phys. Rev. B. {\bf 79}, 224511 (2009). 

\bibitem{ikeda2010} H. Ikeda, R. Arita, and J. Kune\v{s}, Phys. Rev. B
  {\bf 82}, 024508 (2010).

\bibitem{arita} R. Arita and H. Ikeda, J. Phys. Soc. Japan {\bf 78},   
  113707 (2009)

\bibitem{ikeda} H. Ikeda, R. Arita, and J. Kune\v{s}, Phys. Rev. B {\bf 81},
  054502 (2010). 

\bibitem{knolle2011} J. Knolle, I. Eremin, and R. Moessner, Phys. Rev. B {\bf 83}, 224503 (2011).

\bibitem{brydonArXiv}P. M. R. Brydon, J. Schmiedt, and C. Timm, Phys. Rev. B
  {\bf 84}, 214510 (2011). 

\bibitem{kuroki} K. Kuroki, X. Onari, R. Arita, H. Usui, Y. Tanaka,
  H. Kontani, and H. Aoki, Phys. Rev. Lett. {\bf 101}, 087004 (2008). 

\bibitem{graser} S. Graser, T. A. Maier, P. J. Hirschfeld, and D. J. Scalapino,
  New J. Phys. {\bf 11}, 025016 (2009). 

\bibitem{Luo} Q. Luo, G. Martins, D.-X. Yao, M. Daghofer, R. Yu,
  A. Moreo, and E. Dagotto, Phys. Rev. B {\bf 82}, 104508 (2011).

\bibitem{raghu} S. Raghu, Z.-L. Qi, C.-X. Liu, D. J. Scalapino, and
  S.-C. Zhang, Phys. Rev. B {\bf 77}, 220503(R) (2008). 

\bibitem{calderon} M. J. Calder\'{o}n, B. Valenzuela, and E. Bascones,
  Phys. Rev. B {\bf 80}, 094531, (2009). 

\bibitem{miyake} T. Miyake, K. Nakamura, R. Arita, and M. Imada,
  J. Phys. Soc. Jpn. {\bf 79}, 044705 (2010). 

\bibitem{yanagi} Y. Yanagi, Y. Yamakawa, and Y. Ono, J. Phys. Soc. Jpn. {\bf
  77}, 123701 (2008). 

\bibitem{Schickling}T. Schickling, F. Gebhard, J. B\"{u}nemann,
  L. Boeri, O. K. Andersen, and W. Weber, Phys. Rev. Lett. {\bf 108}, 036406
  (2012).

\bibitem{boeri}L. Boeri, O. V. Dolgov, and A. A. Golubov,
  Phys. Rev. Lett. {\bf 101}, 026403 (2008).

\bibitem{vildosola} V. Vildosola, L. Pourovskii, R. Arita, S. Biermann,
  and A. Georges, Phys. Rev. B {\bf 78}, 064518 (2008).

\bibitem{kaneshita} E. Kaneshita, T. Morinari, and T. Tohyama,
  Phys. Rev. Lett. {\bf 103}, 247202 (2009). 

\bibitem{bascones} E. Bascones, M. J. Calder\'{o}n, and B. Valenzuela,
  Phys. Rev. Lett. {\bf 104}, 227201 (2010). 

\bibitem{oles} A. M. Ole\'{s}, Phys. Rev. B {\bf 28}, 327 (1983). 

\bibitem{dagotto} E. Dagotto, T. Hotta, and A. Moreo, Phys. Rep. {\bf 344}, 1
  (2001). 

\bibitem{xu} G. Xu, W. Ming, Y. Yao, X. Dai, S.-C. Zhang, and Z. Fang,
  Europhys. Lett. {\bf 82}, 67002 (2008). 

\end{thebibliography}
\end{document}